\documentclass[aps,pra,superscriptaddress,floatfix,longbibliography, 12pt]{revtex4-2}

\usepackage{amsmath,amssymb} 

\usepackage{colortbl}
\usepackage{booktabs} 
\usepackage{tabulary}
\usepackage{array,longtable,multirow}
\usepackage{tabularx}

\usepackage{bm} 
\usepackage{graphicx} 
\usepackage{comment} 
\usepackage{braket}

\usepackage{amsthm}
\theoremstyle{definition}

\usepackage{enumitem}
\setlist{noitemsep,leftmargin=*,topsep=0pt,parsep=0pt}

\usepackage[dvipsnames]{xcolor} 
\definecolor{lightgray}{gray}{0.6}
\definecolor{medgray}{gray}{0.4}

\usepackage{hyperref}
\hypersetup{colorlinks=true,
    urlcolor= blue,
    citecolor=blue,
    linkcolor= blue}

\newif\ifptitle
\newif\ifpnumber
\newcounter{para}

\ptitletrue  
\pnumbertrue  

\renewcommand{\vec}[1]{\bm{#1}}

\newcommand{\mytitle}{Positive Conserved Quantities in the Klein-Gordon Equation}

\begin{document}

\title{\mytitle}

\author{Robert Lin}
\affiliation{Department of Mathematics, Harvard University,  Cambridge, MA, 02138, USA}
\email{robertlin@g.harvard.edu}

\date{\today}

\begin{abstract}
We introduce an embedding of the Klein-Gordon equation into a pair of coupled equations that are first-order in time. The existence of such an embedding is based on a positivity property exhibited by the Klein-Gordon equation. These coupled equations provide a more satisfactory reduction of the Klein-Gordon equation to first-order differential equations in time than the Schrodinger equation. Using this embedding, we show that the ``negative probabilities" associated with the Klein-Gordon equation do not need to be resolved by introducing matrices as Dirac did with his eponymous equation.
For the case of the massive Klein-Gordon equation, the coupled equations are equivalent to a forward Schrodinger equation in time and a backward Schrodinger equation in time, respectively, corresponding to a particle and its antiparticle. We show that there are two positive integrals that are conserved (constant in time) in the Klein-Gordon equation and thus provide a concrete resolution of the historical puzzle regarding the previously supposed lack of a probabilistic interpretation for the field governed by the Klein-Gordon equation. A significant consequence is that the Schrodinger equation is given a relativistic formulation, which does not require creation and annihilation operators, i.e. quantum fields. Physically, this corresponds to a theory in which the positive and negative energy parts do not directly interact, hence there will be no annihilation events--for example, particle-antiparticle collisions which do not result in photon emission. Thus, one practical consequence of this relativistically consistent theory is a simple explanation for dark matter. 
\end{abstract}

\maketitle

\section{Introduction}

Negative probability has been of interest ever since the formulation of the Klein-Gordon equation, which exhibited a conserved quantity which is nonpositive. In his 1942 article  \cite{Dirac}, Dirac cited this phenomena as one of the foundational difficulties in the theory of quantum fields. In the intervening years, this difficulty has been largely swept under the rug, by the following device which works when the field is associated with a charged scalar particle.  Namely, one identifies the corresponding conserved quantity as a \textit{charge} density (see, e.g., page 38 of the 1961 textbook by Rose \cite{Rose}), rather than a probability density. Since charge readily admits negative and positive values, the difficulty ``disappears."

This device clearly does not work if the particle does not have a charge. In the absence of particles that exhibit the property of being a scalar and are not charged, it is not obviously necessary to resolve this foundational problem. This changed with the discovery of the Higgs particle in 2013 \cite{ATLAS}, which provided the first example of a spin-0 neutral particle that was not manifestly a spin-0 singlet of a composite species. Hence, the foundational problem is still relevant today \footnote{A somewhat parallel problem, also very interesting, is whether perhaps negative probability is a \textit{feature} rather than a bug of the quantum theory. Dirac's 1942 article \cite{Dirac} contemplated the possibility of introducing an indefinite product on wavefunctions, which was envisioned as a possible way to solve the problem of the divergences in quantum field theory. Feynman returned to this question  in a lecture sometime around 1980, which is recounted in the recollections of \cite{Shor2022early}. In that lecture, Feynman contemplated the possibility of evading Bell's theorem by invoking negative probabilities, though later he replaced this motivation by Dirac's original motivation.}.

In this article we resolve this problem by using an embedding method. Our embedding method is rather unusual in that it invokes a construction of an auxiliary field, which \textit{replaces} the role of the wavefunction being acted on by a Hamiltonian operator of a Schrodinger equation. The key in our approach from a conceptual perspective is to use a \textit{phenomenological}  description of the field satisfying the Klein-Gordon equation. That this phenomenological description ends up revealing underlying forward and backward Schrodinger equations in time, is rather pleasant, yet it is not an assumption but rather a \textit{consequence} of the embedding. We discover that in fact there are two positive integrals that are conserved. This corrects the long-standing misunderstanding that there is no positive probability density associated with Klein-Gordon equation (see, for example, Weinberg's discussion of equation (1.2.68) in his book \cite{Weinberg}, in which he states that ``As Dirac had recognized, the only probability density $\rho$ ... must be proportional to the quantity $\rho = 2 \text{Im} \left(\phi^{\dagger} \partial_t \phi \right)$ and therefore is not necessarily a positive quantity." This statement is shown by our article to be incorrect). In fact, what has been historically thought to correspond to a nonpositive probability density is actually an energy density. This accounts for the known result that the historical $\rho$  results in a difference of squares.

From a foundational perspective, our embedding clarifies the role of the Klein-Gordon equation in the theory of relativistic quantum mechanics as \textit{already} establishing the mathematical existence of particles traveling forward or backward in time. In particular, it is apparent from our establishment of the coupled equations in the massive case that the Klein-Gordon equation already contains the germ of Feynman's intuition that antiparticles are particles traveling backward in time, and that what could hitherto be regarded as primarily belonging to the realm of physical intuition turns out to be a direct mathematical consequence of assuming a positivity condition for the Klein-Gordon equation. One of the subtler aspects of our embedding is that it involves a choice of time-like axis. With respect to \textit{each} such a choice, a different embedding is generated. The corresponding pair of conserved quantities thus depends on the choice of reference frame. One may view such a choice as corresponding to the observer's frame of reference. Thus, experiments may be done in the weakly relativistic regime to ascertain whether the Klein-Gordon equation and the resulting coupled equations, rather than a single Schrodinger equation, are needed to account for experimental results in a boosted frame.

The hallmark of the Klein-Gordon equation would be the verification of conserved probability density in a \textit{boosted} frame. Such an experimental verification would rule out the validity of a single forward Schrodinger equation with relativistic Hamiltonian, and instead indicate that \textit{two} Schrodinger equations, one forward and one backward in time, with relativistic Hamiltonian, are necessary to guarantee invariance of the law of probability conservation under Lorentz boosts.

\section{From the Klein-Gordon Equation to Coupled Equations}

\subsection{The Embedding}
The Klein-Gordon equation for a field $\psi$ of mass $m$ in the absence of interactions or an external potential states that
\begin{equation}
	( -\frac{\hbar^2}{c^2}\partial_{tt} + \hbar^2 (\partial_{xx} + \partial_{yy} + \partial_{yy})) \psi = m^2 c^2 \psi.
\end{equation}
We will write the Klein-Gordon equation more generally as
\begin{equation}
	\label{general}
	\hbar^2 \partial_{tt} \psi = - D D^* \psi,
\end{equation}
where $D^*$ denotes the adjoint of $D$ with respect to a fixed inner product $\langle\cdot,\cdot \rangle$ for the space of allowable field configurations. In other words, $\langle D f, g \rangle = \langle f, D^* g \rangle$ is a defining property for the operator $D^*$.

We will show that the following coupled equations can be constructed out of a massive field $\psi(\vec{r},t)$ which obeys the Klein-Gordon equation, provided $D$ is an invertible operator.
\begin{align}
	\hbar \partial_t \psi &= - D \chi \\
	\hbar \partial_t \chi &= D^* \psi.
\end{align}
Note that we do not impose them by \textit{fiat}; instead, the input data is a given spacetime configuration $\psi(\vec{r},t)$, which satisfies the Klein-Gordon equation.

The starting point of our construction of the coupled equations from a prescribed solution $\psi(\vec{x},t)$ to eqn. \ref{general} is to define  
\begin{equation}
	\label{embedding}
	\chi := - D^{-1} (\hbar \partial_t \psi),
\end{equation}
which automatically results in 
\begin{equation}
	\hbar \partial_t \psi = -D \chi.
\end{equation}
This definition is sensible so long as $D$ is invertible.

We then impose the equation of motion
\begin{equation}
	\label{motion}
		\hbar \partial_t \chi = D^* \psi.
\end{equation}
Combining these two equations by applying the equation of motion to the auxiliary field $\chi$ and then applying $-D$ to both sides, readily yields the Klein-Gordon equation
\begin{equation}
	\hbar^2 \partial_{tt} \psi = - D D^* \psi,
\end{equation}
On the other hand, since we can now write
\begin{equation}
	\psi = (D^*)^{-1} (\hbar \partial_t \chi)
\end{equation}
it follows that 
\begin{equation}
	\hbar^2 \partial_{tt} \chi = - D^* D \chi.
\end{equation}

 \subsection{An Example}
 As an example, for the massive Klein-Gordon equation, one requires that
 \begin{equation}
 	D D^* = m^2 c^4 -  \hbar^2 c^2 (\partial_{xx} +\partial_{yy}+\partial_{zz}).
 \end{equation}
Then one possible choice of $D$ is 
\begin{equation}
	D =  i \sqrt{ m^2 c^4 -  \hbar^2 c^2 (\partial_{xx} +\partial_{yy}+\partial_{zz})}
\end{equation}
which yields 
\begin{equation}
D^* =   -i \sqrt{m^2 c^4 -  \hbar^2 c^2 (\partial_{xx} +\partial_{yy}+\partial_{zz}).}
\end{equation}
Here, the operator inside the square roots is nonnegative, and hence it is meaningful to take the square root, by an application of the spectral calculus. Namely, one may consider the spectral decomposition of a nonnegative self-adjoint operator $U$ as $U = \sum_v \lambda_v \ket{v} \bra{v}$, for discrete spectrum, or $U= \int dv\,  \lambda_v \ket{v} \bra{v}$, for continuous spectrum. Then  $\sqrt{U}:= \sum_v \sqrt{\lambda_v} \ket{v} \bra{v}$ or $\int dv \sqrt{\lambda_v} \ket{v} \bra{v}$. 

We also need to verify that $D$ is invertible. The  \textit{positivity} of the operator inside the square root follows from the fact that $- \partial_{xx} - \partial_{yy}-\partial_{zz}$ is known to be nonnegative and self-adjoint, hence adding $m^2 c^4$ ($m>0$) to a positive multiple of this operator results in a strictly positive operator. The square root of a positive operator, as defined above, remains a positive operator, and so the kernel of $D$ must be trivial. We further observe that the  plane waves $e^{i \vec{p}\cdot \vec{x}/\hbar}$ furnish an eigenbasis of $D$, with eigenvalue $i \sqrt{m^2 c^4 + c^2 p^2}$. On physical grounds, we may regard any spacetime configuration $\psi(\vec{x},t)$ as representable by a sum or integral over plane waves of different momentum. Hence, $\psi(\vec{x},t)$ belongs to the image of $D$, and $D$ is invertible. Thus, the condition that $m>0$ ensures that $D$ is invertible, and hence satisfies the criterion for our embedding via equation \ref{embedding}.
 
As a consequence,
\begin{equation}
	D^* = -D,
\end{equation}
and the coupled equations reduce to
\begin{align}
		\hbar \partial_t \psi &= - D \chi \\
	\hbar \partial_t \chi &= -D \psi.
\end{align}
Replacing $D$ by $iH$, it is thus determined that
\begin{align}
		i\hbar \partial_t \psi &= H \chi \\
	i\hbar \partial_t \chi &= H \psi.
\end{align}
Diagonalizing these equations leads to 
\begin{align}
	i\hbar \partial_t \eta_+ &= +H \eta_+ \\
	i\hbar \partial_t \eta_- &= -H \eta_-
\end{align}
Here $\eta_+ = \psi+\chi$, $\eta_- = \psi-\chi$.

Since the quantities $\eta_{\pm}$ respect forward- and backward-in-time Schrodinger equations, respectively, it follows that $\langle \eta_{\pm},\eta_{\pm} \rangle$ are conserved quantities.

Mathematically, what we have done here is construct two operators $\Pi_{\pm}$ satisfying
\begin{equation}
	\Pi_{\pm} = 1\mp D^{-1}(\hbar \partial_t) = 1 \pm i H^{-1} (\hbar \partial_t),
\end{equation}
such that 
\begin{equation}
	\Pi_{\pm} \psi = \eta_{\pm}
\end{equation}
are, up to a normalizing constant, probability amplitudes with conserved \textit{positive} norm. For the standard Euclidean inner product on $\mathbb{R}^3$,  these conserved positive norms are the square roots of the spatial integrals
\begin{equation}
	\int_{\mathbb{R}^3} d^3 \vec{x} |(\Pi_{\pm} \psi)(\vec{x},t)|^2.
\end{equation}

\section{Discussion}
It remains now to compare our result that 
\begin{equation}
	\int_{\mathbb{R}^3} d^3 \vec{x} |( (1 \pm i H^{-1} (\hbar \partial_t)) \psi)(\vec{x},t)|^2
\end{equation}
are conserved quantitites in time with the historical result that  $\rho =2 \text{Im} (\psi^{\dagger} \partial_t \psi)$ is a nonpositive ``probability density"  \cite{Weinberg}.  Note that $2\partial_t \psi = \partial_t \eta_+ + \partial_t \eta_-  =-i( H \eta_+ - H \eta_- )/\hbar$. Multiplying by $\eta_+^* + \eta_-^*$ results in 
\begin{equation}
	\label{express}
	- \frac{i}{\hbar} (\eta_+^* H \eta_+ - \eta_-^* H \eta_-) -\frac{i}{\hbar} \left(\eta_-^* H \eta_+ - \eta_+^* H \eta_-\right)
\end{equation}
Using the self-adjointness of $H$, it follows that the second half of the expression in formula \ref{express} integrates to a \textit{real} number, hence it vanishes when we take the imaginary part. So
\begin{equation}
	\int_{\mathbb{R}^3} d^3 \vec{x} \,\rho(\vec{x},t)  \propto - \frac{2}{\hbar} \int_{\mathbb{R}^3} d^3 \vec{x} (\eta_+^* H \eta_+ - \eta_-^* H \eta_-).
\end{equation}
Evidently, this is the integral of the energy density, up to an overall multiplicative factor. So up to an additive term whose integral vanishes, $\rho$ is proportional to the energy density. It is no wonder it is nonpositive, as it assigns \textit{negative} energy to the forward component in time.

\section{Conclusion}

In summary, the present paper addresses a concrete problem, that of the proper resolution of “negative probabilities” in the Klein-Gordon equation. Dirac attributed great importance to this problem, viewing it as foundational to quantum mechanics. In the present paper, the view is advocated that the Klein-Gordon equation gives rise to two Schrodinger equations, and furthermore we give a precise mechanism by which this occurs. The mechanism preserves relativistic covariance under Lorentz boosts, and conserves positive probabilities. 

This analysis of the Klein-Gordon equation supplies a hitherto unknown instance in which relativistic physics does not lead to particle creation or annihilation, as we separately conserve the total positive and negative energy parts (forward and backward in time) of the field $\psi$, with respect to a given time axis $t$. Namely, a massive scalar particle obeying the Klein-Gordon equation which is traveling close to the speed of light need not induce any particle creation events. This is consistent with the theory of special relativity, as one may boost to a frame in which the scalar particle is moving at speeds very small compared to the speed of light, in which one does not expect any particle creation. 

Thus, a significant consequence is that the Schrodinger equation is given a relativistic formulation, which does not require creation and annihilation operators, i.e. quantum fields.  Whether such a creation/annihilation-free phenomenon can persist under second quantization, i.e. in the presence of quantum fields that are constructed and construed in the new way in which we have interpreted the single-particle theory as arising from the Klein-Gordon equation, would be an interesting problem to pursue. Here, the intuition is that for instance, in a theory in which the positive and negative energy parts do not directly interact, there will be no annihilation events--for example, particle-antiparticle collisions which do not result in photon emission and so the particles are \textit{dark}.  Thus, one practical consequence of this relativistically consistent theory is a simple explanation for dark matter. 

\section*{Acknowledgments}
 I would like to express my gratitude to the late Professor Roy Glauber of Harvard University for encouragement during the nascent beginnings of this work in the spring of 2018. I wish to thank Professor Shing-Tung Yau (now emeritus professor at Harvard University, and professor at Tsinghua University) for indulging my wild ideas regarding the usefulness of embeddings for relativistic quantum mechanics, during a serendipitous conversation, after I had expressed to him my ideas over e-mail, at the time. I thank   Professor Howard Georgi of Harvard University and Professor Stephen Adler of the Institute for Advanced Study for their thoughts at an early stage. I wish to thank Professor Peter Shor of M.I.T. and again Professor Stephen Adler, as well as Professor Lawrence Horwitz of Tel-Aviv University and Bar-Ilan University, for  feedback on the initial preprint of my paper.

\newpage



\end{document}